\documentclass[twoside,7pt]{amsart}
\usepackage{amsmath,amssymb,mathrsfs}
\def\d{\operatorname{d}}\def\<{\langle}\def\>{\rangle}
\def\Tr{\operatorname{Tr}}\def\:{\hbox{\bf :}}
\def\spc#1{\mathcal{#1}}
\def\openone{1\!\!1}
\def\Cmplx{\mathbb C}
\def\Bndd#1{\alg{B}(#1)}\def\map#1{{\mathscr{#1}}}

\def\set#1{{\sf #1}}\def\alg#1{{\mathcal #1}}
\def\grp#1{{\mathbf #1}}

\def\rnk{\operatorname{rank}}
\def\Rng{\set{Rng}}
\def\Supp{\set{Supp}}
\def\Span{\set{Span}}
\def\Klm{{\mathfrak X}}
\def\dim{\operatorname{dim}}

\def\qed{$\,\blacksquare$\par}

\def\dag{\dagger}
 
\newtheorem{Property}{Property}
\newtheorem{Def}{Definition}
\newtheorem{lemma}{Lemma}
\newtheorem{proposition}{Proposition}
\newtheorem{corollary}{Corollary}
\newtheorem{theorem}{Theorem}

\def\>{\rangle}
\def\<{\langle}

\begin{document}
\title{Extremal covariant measurements}
\author{Giulio Chiribella} \email{chiribella@unipv.it}
\author{Giacomo Mauro D'Ariano}
\email{dariano@unipv.it}
\address{{\em QUIT} Group, http://www.qubit.it, Istituto Nazionale di Fisica della Materia,
Unit\`a di Pavia, Dipartimento di Fisica ``A. Volta'', via Bassi 6,
I-27100 Pavia, Italy, and \\ Department of Electrical and Computer
Engineering, Northwestern University, Evanston, IL  60208}
\date{\today}
\maketitle
\begin{abstract}
  We characterize the extremal points of the convex set of quantum
  measurements that are covariant under a finite-dimensional
  projective representation of a compact group, with action of the
  group on the measurement probability space which is generally
  non-transitive. In this case the POVM density is made of multiple
  orbits of positive operators, and, in the case of extremal
  measurements, we provide a bound for the number of orbits and for
  the rank of POVM elements. Two relevant applications are considered,
  concerning state discrimination with mutually unbiased bases and the
  maximization of the mutual information.
\end{abstract}     
\section{introduction}
A fundamental issue in the theory of quantum information
\cite{Nielsen2000} is the investigation of the ultimate precision
limits for extracting classical information from a quantum
system. Indeed, when the information is encoded on quantum states, its
read-out suffers the intrinsically quantum limitation of
discriminating among nonorthogonal states. One then need to optimize
the discrimination with respect to a given optimality criterion, which
is dictated by the particular task for which the measurement is
designed, or by the particular way the information is encoded over
states.  The good news is that, although the position of the problem
has a limited generality due to the specific form of the optimality
criterion, nevertheless for a large class of criteria the optimization
method is given by a standard procedure. In such approach all possible
measurements form a convex set (the convex combination of two
measurements corresponding to the random choice between their
apparatuses), and the optimization corresponds to maximizing a convex
functional---e.~g. the mutual
information\cite{HolevoMutInf,Peres}---or to minimizing a concave
functional--- e.~g. a Bayes cost\cite{Helstrom,Holevo}---over the
convex set of measurements. Since the global maximum of a convex
functional (or the minimum of a concave functional) is achieved over
extremal points, the optimization can be restricted to the extremal
elements of the set only.

In most situations of interest, the set of signal states on which the
information is encoded is invariant under the unitary action of some
group of physical transformations. The symmetry of the set of signal
states then reflects into a symmetry of the optimal measurements,
which without loss of generality can be assumed to be \emph{covariant}
\cite{Holevo} with respect to the same group of transformations.

The problem of charactering extremal covariant measurements has been addressed in Refs.
\cite{ExtPovmAndQo,ExtPovms}, however restricting the analysis to the case of group-action that is
\emph{transitive} on the probability space of measurement outcomes, namely any two points in the
probability space are connected by some group element.  The present
paper completes the investigation by generalizing all results to the
case of non-transitive group actions. Indeed the discrimination of
states belonging to disjoint group orbits occurs in actual
applications, and this situation has received little attention in the
literature. Moreover, when classical information is encoded on quantum
states it can be convenient to decode it with a measurement having
outcomes that are not in one-to-one correspondence with the encoding
states.  This typically happens when the optimality criterion is nonlinear in the probabilities of measurement outcomes, as in the case of the mutual information\cite{shor}.  In the presence of group symmetry, as recently noted
by Decker \cite{Decker}, even if the encoding states form a single
group orbit, the maximization of the mutual information often selects
covariant measurements with probability space that splits into
disjoint orbits. It is then interesting to quantify the number of
orbits needed for the maximization of the mutual information, or at
least to give an upper bound for it. Indeed, as we will see in the
present paper, the characterization of extremal covariant measurements
also provides as a byproduct an alternative and simpler derivation of
the bound given in \cite{Decker}.
\section{Statement of the problem}
In the general framework of quantum mechanics the state of a system is
represented by a density operator $\rho$ on a given Hilbert space
$\spc H$, whereas the statistics of a measurement is described by a
positive operator valued measure (POVM), which associates a positive
semidefinite operator $P(B) \in \Bndd{\spc H}$ to any subset $B\in
\sigma(\Klm)$ of the $\sigma$-algebra of events in the probability space $\Klm$.
The defining properties for a POVM are:
\begin{eqnarray}
&&0 \leq P(B) \leq \openone \qquad \forall B \in \sigma(\Klm)\\
&&P(\cup_{k=1}^{\infty} B_k)= \sum_{k=1}^{\infty} P(B_k)\quad \forall
\{B_k\}~~ \text{disjoint}\\ &&P(\Klm)=\openone~.
\end{eqnarray}
The probability of the event $B \in \sigma(\Klm)$ is then given by the Born rule
\begin{equation}
p(B)= \Tr[\rho P(B)].
\end{equation}

In this paper we will consider the case where the probability space
$\Klm$ supports the action of a compact group $\grp G$, namely any
group element $g \in \grp G$ acts as a measurable automorphism of the
probability space $\Klm$, which maps $\quad x \in \Klm$ to $gx \in
\Klm$. If any two points $x_1, x_2 \in \Klm$ are connected by some
group element---i.e. $x_2= g x_1$ for some $g \in \grp G$---, the
group action is called \emph{transitive}. In this case, which is the
most studied in the literature\cite{Helstrom,Holevo}, the whole
probability space is the group orbit of an arbitrary point $x_0
\in \Klm$, namely $\Klm = \{g x_0~|~ g \in \grp G\}$. In this paper we
will study the more general case where the group action is not
transitive, and, accordingly, the probability space is not a single
group orbit, but the union of a set of disjoint orbits, each one
being labeled by an index $i \in \set I$ for some set $\set I$.  For
simplicity, we will assume the index set $\set I$ to be finite.

The simplest case of non-transitive group action then arises when the
probability space is the Cartesian product of the index set $\set I$
with the compact group $\grp G$, i.e. $\Klm= \set I \times \grp
G$. In this case, the action of a group element $h \in \grp G$ on a point
$x=(i,g)\in \set I
\times \grp G$ is given by $h x =(i,hg)$.
Measurements with outcomes in $\set I \times \grp G$ naturally arise
in the discrimination of a set of signal states which is the union of
a certain number of disjoint group orbits, each orbit $\mathcal{O}_i$
being generated by the action of the group on a given initial state
$\rho_i$, namely $\mathcal{O}_i= \{ U_g \rho_i U_g^{\dag}~|~ g \in
\grp G\}$ for some unitary representation $\set R(\grp G)=\{ U_g ~|~ g \in \grp G\}$. Precisely, if the \emph{stability group} 
$\grp G_i= \{ h \in \grp G~|~ U_h \rho_i
U_h^{\dag}= \rho_i
\}
$ associated to any state $\rho_i$ consists only of the identity element $e$, then there is a one-to-one correspondence between signal states and points of the probability space $\Klm= \set I \otimes \grp G$. In Section 4 we will study in detail the case of POVMs with probability space $\Klm = \set I \times \grp G$. 

If the stability groups associated to the inital states $\{\rho_i~|~i \in \set I\}$ are nontrivial,
namely $\grp G_i \not = \{e\}$ for some $i \in \set I$, in order to have a one-to-one correspondence
between signal states and measurement outcomes, one must consider the probability space $\Klm =
\cup_{i \in \set I} \grp G / \grp G_i$, where $\grp G / \grp G_i$ denotes the quotient of $\grp G$
with respect to the equivalence relation ``$g \sim g'$ if $g'=g \cdot h$ for some $h \in \grp
G_i$''. This more general case will be treated in Section 5.
        
\begin{Def}[covariant POVMs]
Let  $\Klm$ be a probability space supporting the group action $g: x \in \Klm
\mapsto gx \in \Klm$.  A POVM is \emph{covariant}\cite{Holevo} if it satisfies  the property
\begin{equation}
P(B)= U_g^{\dag} P(gB) U_g \qquad \forall B\in \sigma(\Klm),~ \forall
g\in\grp G~,
\end{equation}
where $gB\doteq \{gx ~|~ x\in B\}$.
\end{Def} 
In the case $\Klm = \set I \times
\grp G$, it is simple to prove\cite{proof} that any covariant POVM admits an
operator density $M(i,g)$ with respect to the (normalized) Haar measure $\d g$ on
the group $\grp G$, namely, if $B = (i, A)$, where $A \subseteq \grp G$ is a measurable subset, then $P(B)= \int_A~ \d g~ M(i,g)$.
Moreover, such an operator density has necessarily the form \cite{proof}  
\begin{equation}\label{POVMDens}
M(i,g)= U_g~ A_i~ U_g^{\dag}~,
\end{equation}
where $A_i \in \Bndd{\spc H}$ are Hermitian operators satisfying the constraints
\begin{equation}\label{PosSeed}
A_i \ge 0 \qquad \forall i \in \set I
\end{equation}
\begin{equation}\label{NormCovPovm}
\sum_{i \in \set I} \int_{\grp G} \d g~ U_g~A_i~ U_g^{\dag}= \openone ~. 
\end{equation}
Here and throughout the paper we adopt for the Haar measure the normalization  
\begin{equation}
\int_{\grp G} \d g=1~.
\end{equation} 

According to the above discussion, any covariant POVM with probability space $\Klm = \set I \otimes \grp G$ is completely
specified by a set of operators $\{A_i~|~i \in \set I\}$, such that
both constraints in Eqs. (\ref{PosSeed}) and (\ref{NormCovPovm}) are
satisfied.  Moreover, it turns out that it is very useful to represent
such a vector of operators as a single block operator $A= \bigoplus_{i
\in \set I}~ A_i~$, acting on an auxiliary Hilbert space $\spc H_{aux}
\doteq \bigoplus_{i \in \set I}
\spc W_i$, where $\spc W_i \simeq \spc H \quad \forall i \in \set
I$. In terms of the block operator $A \in \bigoplus_{i \in \set I}
\Bndd {\spc W_i}$ the two constraints Eq. (\ref{PosSeed}) and
Eq. (\ref{NormCovPovm}) become
\begin{equation}\label{BlockPos}
A \ge 0
\end{equation}
and 
\begin{equation}\label{LinAppl}
\map L (A) = \openone~,
\end{equation}
where $\map L: \bigoplus_{i \in \set I}
\Bndd {\spc W_i} \to \Bndd {\spc H}$ is the linear map 
\begin{equation}\label{LA}
\map L(A)\doteq \sum_{i \in \set I} \int_{\grp G} \d g~ U_g A_i U_g^{\dag}~.
\end{equation}
The two constraints (\ref{BlockPos}) and (\ref{LinAppl}) define such a convex subset of the space of block operators $\bigoplus_{i \in \set I} \Bndd {\spc W_i}$, which is in one-to-one affine correspondence with the convex set of covariant POVMs. In the following, the convex set of block operators will be denoted by $\set C$.
\begin{proposition}
The convex set $\set C$, defined by the constraints (\ref{BlockPos}) and (\ref{LinAppl}) is compact in the operator norm.
\end{proposition}
\proof Since $\set C$ is a subset of a finite dimensional vector space, it enough to show that $\set
C$ is bounded and closed. $\set C$ is bounded, since for any $A \in \set C$, one has $||A|| \le \Tr[A] =\sum_{i \in \set I} \Tr[A_i]=\Tr[\map L(A)] =d$ (using Eqs. (\ref{BlockPos}) and (\ref{LinAppl})).  Moreover, $\set C$ is
closed. In fact, if $\{A_n\}$ is a Cauchy sequence of points in $\set C$, then $A_n$ converges to
some block operator $A \in \bigoplus_{i \in \set I} \Bndd {\spc W_i}$.  We claim that $A$ belongs to
$\set C$. Of course, $A$ satisfies condition (\ref{BlockPos}). As regards condition (\ref{LinAppl}),
just notice that the $\map L$ is continuous, being linear. Therefore, we have $||\map L(A)
-\openone||= ||\map L (A-A_n) || \to 0$, namely $A$ satisfies condition (\ref{LinAppl}).\qed 
\emph{Observation 1.} Since the convex set $\set C$ is compact, it
coincides with the convex hull of its extreme points, i.e. any element
$A \in \set C$ can be written as convex combination of extreme points.
The classification of the extreme points of $\set C$ will be given in
Section 4.

\emph{Observation 2.}
In this section and all throughout the paper, $\grp G$ is assumed to
be a compact Lie group. Nevertheless, all results clearly hold also if
$\grp G$ is a finite group, with cardinality $|\grp G|$. In this case, one only has to make the
substitution $\int_{\grp G} \d g~
\to ~ \frac 1 {|\grp G|} \sum_{g \in \grp G}$. Moreover, since now the probability space $\Klm =
\set I \times \grp G$ is discrete, there is no need of introducing any
operator density, and we simply have 
\begin{equation}
P(i,g)=\frac 1 {|\grp G|}~
U_g A_i U_g^{\dag}~.
\end{equation}
An example of covariant POVM with a finite symmetry group will be given in Section 6.

\section{Some results of elementary group theory}\label{GrpTools}
Let be $\grp G$ a compact Lie group and let be $\d g$ the invariant
Haar measure on $\grp G$, normalized such that $\int_{\grp G} \d g=1$.
Consider a finite dimensional Hilbert space $\spc H$ and represent
$\grp G$ on $\spc H$ by a unitary (generally projective) representation $\set
R (\grp G)=\{U_g~|~ g\in\grp G\}$. The collection of equivalence
classes of irreducible representations which show up in the
decomposition of $\set R (\grp G)$ will be denoted by $\set S$. Then
$\spc H$ can be decomposed into the direct sum of orthogonal irreducible
subspaces:
\begin{equation}
\spc H = \bigoplus_{\mu \in S} \bigoplus_{k=1}^{m_{\mu}} \spc H_k^{\mu}~,
\end{equation}
where the index $\mu$ labels equivalence classes of irreducible
representations (irreps), while the index $i$ is a degeneracy
index labeling $m_{\mu}$ different equivalent representations in the
class $\mu$.  Subspaces carrying equivalent irreps have all
the same dimension $d_{\mu}$ and are connected by invariant
isomorphisms, namely for any $k,l=1, \dots, m_{\mu}$ there is an
operator $T_{kl}^{\mu} \in \Bndd {\spc H}$ such that $\Supp (
{T_{kl}^{\mu}})= \spc H_l^{\mu}$, $\Rng ( T_{kl}^{\mu}) = \spc
H_k^{\mu}$, and $[T_{kl}^{\mu}, U_g]=0 \quad
\forall g \in
\grp G$. 
Due to Schur lemmas, any operator $O$ in the commutant of the representation $\set R (\grp G)$ has the form:
\begin{equation}\label{commutingO}
O=\sum_{\mu} \sum_{k,l=1}^{m_{\mu}}~~ \frac{\Tr[T_{lk}^{\mu} O]}{d_{\mu}}~ T_{kl}^{\mu}~.
\end{equation}  
Using the above formula, the normalization of a covariant POVM, given
by Eq. (\ref{LinAppl}), can be rewritten in a simple form. In
fact, due to the invariance of the Haar measure $\d g$, we have $[\map L(A),U_g]=0 \quad \forall g \in \grp G$, i.e. $\map L(A)$ belongs to the commutant of
$\set R (\grp G)$. Then, by exploiting Eq. (\ref{commutingO}), we rewrite the normalization constraint (\ref{LinAppl}) as
\begin{equation}\label{TrNorm}
\sum_{i \in \set I}  \Tr[T_{kl}^{\mu}~ A_i]=  d_{\mu} ~\delta_{kl}\qquad \forall \mu \in S,\quad \forall k,l=1,\dots,m_{\mu}~,
\end{equation} 
$\delta_{kl}$ denoting the Kronecker delta.

Again, this condition can be recasted into a compact form by introducing the
auxiliary Hilbert space $\spc H_{aux} = \bigoplus_{i \in \set I} \spc
W_i$, with $\spc W_i \simeq \spc H \quad \forall i \in \set I$, and constructing a block operator
with a repeated direct sum of the same operator $T_{kl}^{\mu}$, i. e. 
\begin{equation}
S_{kl}^{\mu} = \bigoplus_{i \in \set I} S_{kli}^{\mu}~,\qquad S_{kli}^{\mu} = T^{kl}_{\mu} \quad
\forall i \in \set I. 
\end{equation}
With this definition, Eq. (\ref{TrNorm}) becomes
\begin{equation}\label{CompactTrNorm}
\Tr[ S_{kl}^{\mu}~A]= d_{\mu} \delta_{kl}~, 
\qquad \forall \mu \in S,\quad \forall k,l=1,\dots,m_{\mu}~,
\end{equation} 
where $A$ is the block operator $A = \bigoplus_{i \in \set I}~ A_i$. 

\section{Extremal covariant POVMs}\label{Sec:ExtCovPovm}
This section contains the main result of the paper, namely the
characterization of the extremal covariant POVMs with probability
space $\set I \otimes \grp G$.  Such a characterization will be given by exploiting the one-to-one affine corrispondence between the convex set of covariant POVMs and the convex set $\set
C$ of block operators defined by the constraints (\ref{BlockPos}) and
(\ref{LinAppl}), or, equivalently, by (\ref{BlockPos}) and (\ref{CompactTrNorm}).

\begin{Def} An Hermitian block operator $P = \bigoplus_{i \in
\set I}~P_i$ is a perturbation of $A \in \set C$
if there exists an $\epsilon > 0$ such that $A + t P \in \set C$ for
any $t \in [-\epsilon, \epsilon]$.
\end{Def}
Clearly, a point $A \in \set C$ is extreme if and only if it admits
only the trivial perturbation $P=0$.
\begin{lemma}\label{Lemma:Pert}
A block operator $P= \bigoplus_{i \in \set I}~ P_i$ is a perturbation of $A \in \set C$ if and only if
\begin{eqnarray}
\label{Supp} &&\Supp (P) \subseteq \Supp (A)\\ 
\label{Tr0} && \Tr[ S^{\mu}_{kl}~P] =0  \qquad  \forall \mu \in S,\quad \forall k,l=1,\dots,m_{\mu}
\end{eqnarray}
\end{lemma}
\proof 
Condition (\ref{Supp}) is equivalent to the existence of an $\epsilon >0$ such that $A + t P \ge 0$ for all $t \in
[-\epsilon,\epsilon]$ (see Lemma 1 of
Ref. \cite{ExtPovms}). On the other hand, condition (\ref{Tr0}) is equivalent to require that $A + t P$
satisfies the normalization constraint (\ref{TrNorm}) for all $t \in [-\epsilon,\epsilon]$. \qed
\emph{Observation.} Note that, due to the block form of both $P$ and $A$, condition (\ref{Supp}) is equivalent to 
\begin{equation}
\Supp (P_i) \subseteq \Supp (A_i) \qquad \forall i \in \set{I}~. 
\end{equation}

Using the previous lemma, we can obtain a first characterization of extremality:
\begin{theorem}[Minimal support condition]
A point $A \in \set C$ is extremal if and only if for any $B \in \set C$,
\begin{equation}\label{MinSuppCond}
\Supp(B) \subseteq \Supp (A)~ \Longrightarrow ~ A=B~. 
\end{equation}
\end{theorem}
\proof 
Suppose $A$ extremal. Then, if $\Supp (B) \subseteq \Supp (A)$,
according to Lemma \ref{Lemma:Pert}, $P= A-B$ is a perturbation of
$A \in \set C$. Then $P$ must be zero. Viceversa, if $P$ is a
perturbation of $A$, then $B= A +t P$ is an element of $\set C$ for
some $t \not =0$. Due to Lemma \ref{Lemma:Pert}, we have $\Supp (B)
\subseteq \Supp (A)$. Then, condition (\ref{MinSuppCond}) implies $B= A + t
P=A$, i.e. $P=0$. Therefore, $A$ is extremal. \qed

\begin{corollary}\label{cor:RankOneExt}
If $A \in \set C$ and $\rnk (A)=1$, then $A$ is extremal.
\end{corollary}
\proof Since $\rnk (A)=1$, then, for any $B \in \set C$, the condition $\Supp (B) \subseteq \Supp (A)$  
 implies $B = \lambda A$ for some $\lambda > 0$. Moreover,
 since both $A$ and  $B$ are in $\set C$, from Eq. (\ref{CompactTrNorm}) we have $d_{\mu} =
 \Tr[S_{kk}^{\mu} B] = \lambda \Tr[S_{kk}^{\mu} A] = \lambda d_{\mu}$,
 whence necessarily $\lambda =1$. Condition (\ref{MinSuppCond}) then ensures that $A$ is extremal. \qed 

A deeper characterization of extremal covariant POVMs can be obtained
by using the following lemma.

\begin{lemma}\label{Lemma:Pert1} 
Let $A$ be a point of $\set C$, represented as 
\begin{equation}
A= \bigoplus_{i \in \set I}~ X_i^{\dag} X_i~,
\end{equation}
and define $\spc H_i= \Rng (X_i)$ the range of $X_i$. 
A block operator $P= \bigoplus_{i \in \set I} P_i$ is a perturbation of $A$ if and only if 
\begin{equation}\label{XForm}
P_i= X_i^{\dag}~Q_i~X_i \qquad \forall i \in \set I ~,
\end{equation}
for some Hermitian $Q_i \in \Bndd{\spc H_i}$, and 
\begin{equation}\label{Tr0Bis}
\sum_{i \in \set I}~ \Tr[ S_{kli}^{\mu}~ X_i^{\dag} Q_i X_i ]=0~.
\end{equation}

\end{lemma}
\proof 
First of all, the form (\ref{XForm}) is equivalent to condition
(\ref{Supp}). In fact, if $P$ has the form (\ref{XForm}), then clearly
$\Supp (P)
\subseteq \Supp (A)$. Viceversa, if we assume condition (\ref{Supp}) and write $P=
\bigoplus_{i \in \set I}~ P_i$, we have necessarily $\Supp(P_i) \subseteq \Supp (X_i^{\dag} X_i) = \Supp (X_i)$. Exploiting the singular value decomposition
$X_i=\sum_{n=1}^{r_i} \lambda_{n}^{(i)}
|w_n^{i}\>\<v_n^{i}|$, where $\{|v_n^{i}\>\}$ and
$\{|w_n^{i}\>\}$ are orthonormal bases for $\Supp(X_i)$ and
$\Rng(X_i)$ respectively, we have that any Hermitian operator $P_i$ satisfying $\Supp(P_i) \subseteq
\Supp(X_i)$ has the form $P_i~=~\sum_{m,n} p^{(i)}_{mn} |v_m\>\<v_n|$, whence it can be written as $P_i=X_i^{\dag}Q_i X_i$, for some suitable Hermitian operator
$Q_i \in \Bndd {\Rng (X)}$. Once the equivalence between the form (\ref{XForm}) and condition (\ref{Supp}) is established, relation (\ref{Tr0Bis}) follows directly from Eq. (\ref{Tr0}). \qed

\emph{Observation:} 
According to the previous lemma, a perturbation of $A$ is completely
specified by a set of Hermitian operators $\{Q_i \in {\Bndd{\spc H_i}}~|~ i \in
\set{I}\}$, where $\spc H_i = \Rng (X_i)$. Such operators can be casted into a single block operator
$Q \in \bigoplus_{i \in \set I} \Bndd{\spc H_i}$ by defining
\begin{equation}
Q = \bigoplus_{i \in \set I} Q_i~.
\end{equation}
In terms of the block operator $Q$ we have the following:
\begin{lemma}\label{Lemma:PertBis}
Let  $A = \bigoplus_{ i \in \set I} X_i^{\dag} X_i$ be a point of $\set C$. 
Define the block operators 
\begin{equation}\label{F}
F_{kl}^{\mu} = \bigoplus_{i \in \set I} X_i~ S^{\mu}_{kli}  X_i^{\dag}~.
\end{equation}
Then $A$ admits a perturbation if and only if there exists an Hermitian block operator  $Q\in \bigoplus_{i \in \set I} \Bndd {\spc H_i}$ such that 
\begin{equation}\label{Tr0F}
\Tr[ F_{kl}^{\mu} Q]=0~, \qquad \forall  \mu \in \set S,~ \forall k,l =1, \dots , m_{\mu}~.
\end{equation} 
\end{lemma}
\proof 
Using the definition of $F^{\mu}_{kl}$ and the cyclic property of the trace, it is immediate to see the Eq. (\ref{Tr0F}) is
equivalent to Eq. (\ref{Tr0Bis}). \qed

The previous lemma enables us to characterize the extremal points of
$\set C$.
\begin{theorem}[Spanning set condition]
Let be $A= \bigoplus_{i \in \set I}~ X_i^{\dag} X_i$ be a point of
$\set C$, and $\set F =\{ F_{kl}^{\mu}~|~ \mu \in \set S, k,l=1,
\dots, m_{\mu}\} $ be the set of block operators defined in Lemma
\ref{Lemma:PertBis}.  Then, $A$ is extremal if and only if
\begin{equation}\label{SpanSetCond}
\Span (\set F) = \bigoplus_{i \in \set I} \Bndd{\spc H_i}~,
\end{equation}
where $\spc H_i = \Rng (X_i)$.
\end{theorem}
\proof
$A$ is extremal iff it admits only the trivial perturbation $P=0$. Equivalently, due
to Lemma \ref{Lemma:PertBis}, $A$ is extremal iff the only Hermitian
operator $Q \in \bigoplus_{i \in \set I} \Bndd {\spc H_i}$ that
satisfies Eq. (\ref{Tr0F}) is the null operator $Q=0$. Let us decompose the
Hilbert space $\spc K = \bigoplus_{ i \in \set I} \Bndd {\spc H_i}$,
as $\spc K = \Span (\set F) \oplus \Span (\set F)^{\perp}$,
where $\perp$ denotes the orthogonal complement with respect to the
Hilbert-Schmidt product $(A,B) = \Tr[A^{\dag}B]$. Then, $A$ is extremal iff the only Hermitian operator in $\Span (\set F)^{\perp}$ is the null operator. This is equivalent to the condition $\Span (\set F)^{\perp}=\{0\}$, i.e. $\spc K = \Span (\set F)$. \qed
\begin{corollary}\label{Cor:BoundRank}
Let $A=\bigoplus_{i \in \set I} X_i^{\dag} X_i$ be a point of $\set C$,
and let define $r_i = \rnk X_i$. If $A$ is
extremal, then the following relation holds
\begin{equation}\label{BoundRank}
\sum_{i \in \set I}ì r_i^2 \le \sum_{ \mu \in \set S}~ m_{\mu}^2~. 
\end{equation}
\end{corollary}
\proof
For an extreme point of $\set C$, relation (\ref{SpanSetCond}) implies
that the cardinality of the set $\set F$ is greater than the
dimension of $\spc K = \bigoplus_{i \in \set I} \Bndd {\spc H_i}$.
Then, the upper bound (\ref{BoundRank}) follows from $\dim \spc K = \sum_{i \in
\set I} r_i^2$ and from the fact that  $|\set F|= \sum_{\mu \in \set S} m_{\mu}^{2}$. \qed
\emph{Observation.} 
If the group-representation $\set R (\grp G)$ is irreducible, than its
Clebsch-Gordan decomposition contains only one term $\bar \mu$ with
multiplicity $m_{\bar \mu}=1$. Then, bound (\ref{BoundRank}) becomes
$\sum_{i \in \set I} r_i^2 \le 1$, namely for an extremal $A=\bigoplus_{i \in \set I} A_i$, one has
necessarily $\rnk (A_{i_0})=1$ for some $i_0 \in \set I$, and
$A_i =0$ for any $i \not = i_0$ (this is also a sufficient
condition, due to Corollary \ref{cor:RankOneExt}).  In terms of the
corresponding covariant POVM $M(i,g)= U_g~A_i~U_g^{\dag}$, one has
$M(i,g)=0$ for any $i \not = i_0$, i.e. corresponding to events in the probability space that never occur.
 
\section{Extremal covariant POVMs in the presence of nontrivial stability groups}
In the previous section, we obtained a characterization of extremal
covariant POVMs whose probability space is $\Klm = \set I \times \grp
G$ for some finite index set $\set I$. The framework we outlined is
suitable for a straightforward generalization to the case $\Klm =
\cup_{i \in \set I} \grp G / \grp G_i$, where $\grp G_i$ are compact
subgroups of $\grp G$.  

In this case, it is possible to show that a covariant POVM $P$ admits
a density $M(x_i)$ such that for any measurable subset $B\subseteq
\grp G /
\grp G_i$ one has $P(B) \equiv P_i (B)\doteq \int_{B_i} \d x_i M(x_i)$, where $\d x_i$ is the group invariant measure on $\grp G /\grp G_i$. The form of the operator density is now
\begin{equation}\label{OpDensStabGroup}
M(x_i) = U^{\phantom{\dag}}_{g_i(x_i)}~A_i~U^{\dag}_{g_i(x_i)}~,
\end{equation}
where $A_i \ge 0$, and $g_i(x_i) \in \grp G$ is any representative element of the
equivalence class $x_i\in\grp G / \grp G_i$. The normalization of the POVM is still given
by Eq. (\ref{TrNorm}). In addition, in order to remove the dependence of $M(x_i)$ from the
choice of the representative $g_i(x_i)$, each operator $A_i$ must satisfy the relation
\begin{equation}\label{Comm}
[A_i, U_h]=0 \quad \forall h \in \grp G_i~.
\end{equation}
The commutation constraint (\ref{Comm}) can be simplified by
decomposing each representation $\set R (\grp G_i)= \{U_h~|~ h \in \grp G_i\}$ into
irreps
\begin{equation}
U_h=\bigoplus_{\nu \in \set S_i}
U_h^{\nu_i}~\otimes~\openone_{m_{\nu_i}}~,
\end{equation}
where $m_{\nu_i}$ denotes the multiplicity of the irrep $\nu_i$, and $\set S_i$ denotes
the collection of all irreps contained in the decomposition of $\set R (\grp G_i)$. 
This corresponds to the decomposition of the Hilbert space $\spc H$ as
\begin{equation}
\spc H =\bigoplus_{ \nu_i \in \set S_i}  \spc H_{\nu_i} \otimes \Cmplx^{m_{\nu_i}}~, 
\end{equation} 
where $\spc H_{\nu_i}$ is a representation space, supporting the
irrep $\nu_i$, and
$\Cmplx^{m_{\nu_i}}$ is a multiplicity space.  In this decomposition,
the commutation relation (\ref{Comm}) is equivalent to the block form
\begin{equation}
A_i = \bigoplus_{\nu_i \in \set S_i}  \openone_{\nu_i} \otimes A_{i,\nu_i}~,
\end{equation} 
where $A_{i,\nu_i}\ge 0$ are operators acting on the multiplicity space $\Cmplx^{m_{\nu_i}}$.

By defining  $\omega =(i,\nu_i)$ and $\Omega = \cup_{i \in \set I}~ S_i$,  we can introduce an auxiliary Hilbert space, and associate to a covariant POVM the block operator
\begin{equation}
A= \bigoplus_{\omega  \in \Omega}  A_{\omega }~,
\end{equation} 
where $A_{\omega }\doteq A_{i,\nu_i}$. 
Furthermore, we define the block operators
\begin{equation}
S_{kl}^{\mu} = \bigoplus_{\omega  \in \Omega}  S_{kl\omega }^{\mu}~,
\end{equation}
where now $S_{kl\omega }= \Tr_{\spc H_{\nu_i}} [\Pi_{\nu_i}
T_{kl}^{\mu}]$. Here $\Pi_{\nu_i}$ denotes the projector onto $\spc
H_{\nu_i}~\otimes~\Cmplx^{m_{\nu_i}}$, and $\Tr_{\spc H_{\nu_i}}$
denotes the partial trace over $\spc H_{\nu_i}$.  With these definitions, the normalization of the POVM, given by Eq. (\ref{TrNorm}), becomes equivalent to
\begin{equation}\label{CompactTrNorm2}
\Tr[S_{kl}^{\mu}~ A] = \delta_{kl}~ d_{\mu}~.
\end{equation}

Now we call $\set D$ the convex set of block operators $A =
\bigoplus_{\omega  \in \Omega } A_{\omega }$, defined by
the two conditions $A \ge 0$ and Eq. (\ref{CompactTrNorm2}). Such a
convex set is in one-to-one affine correspondence with the convex set
of covariant POVMs with probability space $\Klm = \cup_{i \in \set I}
\grp G /\grp G_i$. Since the constraints defining $\set D$ are
formally the same defining the convex set $\set C$, we can exploit the
characterization of extremal points of the previous section. In
particular, Corollary \ref{Cor:BoundRank} becomes
\begin{corollary}\label{Cor:BoundRank2}
Let $A= \bigoplus_{\omega  \in \Omega } X_{\omega }^{\dag}X_\omega$ 
be a point of $\set D$, and define $r_{i,\nu_i}\equiv r_{\omega}=\rnk
(X_{\omega})$. If $A$ is extremal, then the following relation holds:
\begin{equation}\label{BoundRank2}
\sum_{i \in \set I} \sum_{\nu_i \in \set S_i} r_{i,\nu_i}^2 \le \sum_{\mu \in  \set S} m_{\mu}^2~. 
\end{equation} 
\end{corollary}
\emph{Observation.} As in the case of Corollary \ref{Cor:BoundRank}, if the representation $\set R (\grp G)$ is irreducible, as a consequence of the bound about ranks, one obtains $\rnk (A_{\omega_0})=1$ for some $\omega_0 \in \set \Omega$, and $A_{\omega}=0$ for any $\omega \not = \omega_0$.  
\section{Applications}
Here we give two examples of the use of the characterization of
extremal POVMs in the solution of concrete optimization problems.
\subsection{State discrimination with mutually unbiased bases} 
Here we consider a case of state discrimination where the set of
signal states is the union of two mutually unbiased bases\cite{WoottersFields} related by
Fourier transform. Precisely, let $\spc H$ be a $d$-dimensional
Hilbert space, and consider the orthornormal bases $\mathcal{B}_1
=\{|n\>~|~ n=0, \dots, d-1\}$ and $\mathcal{B}_2=\{|e_n\>~|~n=0,
\dots, d-1\}$, where 
$|e_n\>=\frac 1 {\sqrt{d}}~\sum_{m=0}^{d-1}~
\omega^{mn} |m\>$, $\omega~=~\exp \left(\frac{2 \pi i} d \right).$
$\mathcal{B}_1$ and $\mathcal{B}_2$ are mutually unbiased, namely
$|\<m| e_n\>|^2=1/d$ for any $m,n$. Consider the two sets of states
defined by $\mathcal{S}_1= \{\rho_{1n} = |n\>\<n|~|~n=0, \dots, d-1\}$ and
$\mathcal{S}_2= \{\rho_{2n}=|e_n\>\<e_n|~|~ n=0, \dots, d-1\}$. Now the
problem is to determine with mimimum error probability the state of
the system, which is randomly prepared either in a state of
$\mathcal{S}_1$ with probability $p/d$, or in a state of $\mathcal{S}_2$ 
with probability $(1-p)/d$.

Exploiting the results of the present paper it is immediate to find the
measurement that minimizes the error probability. In fact, let us
consider the irreducible representation of the group $\grp G = \mathbb{Z}_d \times \mathbb{Z}_d$ given by 

\begin{equation}
\set R (\grp G)=\left\{ U_{pq}=\sum_{n=0}^{d-1}~\frac {\omega^{qn}}{\sqrt{d}}~|n~\oplus~p\>\<n|~,~ (p,q) \in  \mathbb{Z}_d \times \mathbb{Z}_d\right\}~,
\end{equation}
where $\oplus$ denotes addition modulo $d$. Then, the sets
$\mathcal{S}_1$ and $\mathcal{S}_2$ are the group orbits of the inital
states $\rho_{10}$ and $\rho_{20}$, respectively. Moreover, the states
$\rho_{10}$ and $\rho_{20}$ have nontrivial stability groups $\grp G_1$ and
$\grp G_2$, defined by the unitaries $\set R (\grp G_1) =\{U_{0q}~|~ q
\in\mathbb{Z}_d\}$ and $\set R( \grp G_2)=\{U_{p0}~|~p\in \mathbb{Z}_d\}$.  Therefore, signal states are in one-to-one correspondence with points of the probability space
$\Klm = \grp G/\grp G_1 ~\cup~ \grp G/\grp G_2$, such points being
denoted by couples $(i,n)$ where $i\in \{1,2\}$ and $n \in
\mathbb{Z}_d$. For the discrimination we can consider without loss of generality a covariant POVM, of the form of
Eq. (\ref{OpDensStabGroup}), where now the group element $g$ is the
couple $(p,q) \in \mathbb{Z}_d \times \mathbb{Z}_d$.  Moreover, since
the probabilities are linear in the POVM, in the minimization of the
error probability we can restrict the attention to extremal covariant
POVMs. Now, the representation $\set R (\grp G)$ is irreducible,
whence Corollary \ref{Cor:BoundRank2} requires either $A_1=0$ or
$A_2=0$ in Eq. (\ref{OpDensStabGroup}). This means that either the
states in $\mathcal{S}_1$ or the states in $\mathcal{S}_2$ are never
detected. Moreover, since the states within a given set, either
$\mathcal{S}_1$ or $\mathcal{S}_2$, are orthogonal, they can be
perfectly distinguished among themselves. Therefore, the optimal POVM
is  $P^{(1)}(i,n) =
\delta_{i1} |n\>\<n|$ if $p \ge 1/2$, and  $P^{(2)}(i,n) = \delta_{i2} |e_n\>\<e_n|$ otherwise. In particular, if $p=1/2$, an experimenter who tries to discriminate states of two Fourier transformed bases cannot do anything better than randomly choosing one of the orthogonal measurements $P^{(1)}$ and $P^{(2)}$.
This is the working principle of the BB84 crypthographic protocol\cite{BB84}.

\emph{Observation.} 
The previous result can be easily generalized to a case of state
discrimination with more than two mutually unbiased bases. In fact, if
we have a set of mutually unbiased bases $\{\mathcal{B}_i~|~ i \in
\set I\}$ that are all generated by the irreducible representation
$\set R(\grp G)=\{U_{pq}~,~ (p,q)
\in  \mathbb{Z}_d \times  \mathbb{Z}_d\}$, all considerations about extremal covariant POVM still hold. If $\mathcal{S}_i$ is the set of states associated to the basis $\mathcal{B}_i$, and $p_i/d$ is the probability of extracting a state from $\mathcal{S}_i$ ($\sum_{i\in \set I} p_i =1$),  then the covariant POVM which discriminates the signal states with minimum error probability is the orthogonal measurement onto the basis $\mathcal{B}_{\bar l}$ such that $p_{\bar l} = \max_{l \in \set I} \{p_l\}$.
Notice that, if the dimension of the Hilbert space $\spc H$ is $d=p^r$, where $p$ is some prime number, then there are $d+1$ MUBs that are generated by the irreducible representation $\set R(\grp G)$ via the construction by Wootters and Fields\cite{WoottersFields}. 

\subsection{Maximization of the mutual information}
A frequent problem in quantum communication is to find the POVM $P_i,~i\in \set I$, that maximizes the mutual information with a
given set of signal states $\mathcal{S}=\{\rho_j~|~ j \in \set
J\}$. Denoting by $p_j$ the probability of the signal state $\rho_j$, by $q_i =
\sum_{j \in \set J} p_j \Tr[M_i \rho_j]$ the  overall probability of the outcome $i$, and by $p_{ij}=p_j \Tr[M_i \rho_j]$ the joint probability of the outcome $j$ with the state $\rho_i$, the mutual information is defined as 
\begin{equation}
I= H(\{p_{ij}\}) - H(\{p_i\}) -H(\{q_j\})~, 
\end{equation} 
where $H(\{p_i\})\doteq \sum_i -p_i \log (p_i)$ is the Shannon
entropy.  As in the minimization of a Bayes cost\cite{Helstrom,Holevo}, when the set of
signal states is invariant under the action of some finite group $\grp
G$ and all states in the same group orbit have the same probability, one
can without loss of generality restrict the search for the optimal
POVM among covariant POVMs with probability space $\Klm = \set I
\otimes \grp G$, for some finite index set $\set
I$\cite{Davies,Decker}. However, differently from the case of state discrimination, the points of the probability space do not need to be
in one-to-one correspondence with the signal states. Therefore, the set
$\set I$ is not specified \emph{a priori}.  

Combining our characterization of extremal covariant POVMs with the following basic properties of
the mutual information (for the proofs, see Ref.\cite{Davies}), we can readily obtain a bound about
the cardinality of the index set $\set I$.
\begin{Property}\label{p1} The mutual information is a convex functional of the POVM.
\end{Property}
\begin{Property}\label{p2} 
In the maximization of the mutual information, one can consider without loss of generality POVMs made of rank-one operators.
\end{Property}
Consider a covariant POVM $P(i,g) =\frac 1 {|\grp G|}~U_g A_i U_g^{\dag}$. Due to Property \ref{p1},
in the maximization of the mutual information we can consider extremal covariant POVMs. Then, from
Corollary \ref{Cor:BoundRank}, we have the bound $\sum_{i \in \set I} \rnk (A_i)^2 \le \sum_{\mu \in
  \set S} m_{\mu}^2$. Due to Property \ref{p2}, this also implies that the number of (rank-one)
operators $A_i$ must be smaller than $\sum_{\mu \in \set S} m_{\mu}^2$. Therefore, we can assume
without loss of generality
\begin{equation}
|~\set I~| \le \sum_{\mu \in \set S} m_{\mu}^2~.
\end{equation} 
This provides an alternative derivation of the bound given in Ref.\cite{Decker}. 
Finally, if the representation $\set R (\grp G)$ is irreducible, the bound gives $|\set I|=1$, namely the probability space is $\Klm \simeq \grp G$, according to the classic result of \cite{Davies}.
\section{Acknoledgements} 
This work has been supported by Ministero Italiano dell'Universit\`a e  
della Ricerca (MIUR) through FIRB (bando 2001) and PRIN 2005.

\end{document}